\documentclass[11pt]{article}

\newfont{\goth}{eufm10 scaled\magstep1}

\let\ssection=\section
\renewcommand{\section}{\setcounter{equation}{0}\ssection}

\def\parag{\hfil\break} 
\def\kikezd{\parag\underbar}
\def\p{{\partial}}

\def\vP{{\vec P}}
\def\vA{{\vec A}}
\def\vQ{{\vec Q}}
\def\vE{{\vec E}}
\def\vX{{\vec X}}
\def\vsigma{{\vec\sigma}}
\def\IR{{\bf R}}

\newcommand\FS{{F\cdot S}}

\begin{document}

\setlength{\baselineskip}{16pt}

\title{Comments on spin-orbit interaction of anyons} 

\author{
P.~A.~Horv\'athy\footnote{e-mail: horvathy@univ-tours.fr}
\\
Laboratoire de Math\'ematiques et de Physique Th\'eorique\\
Universit\'e de Tours\\
Parc de Grandmont\\
F-37 200 TOURS (France)
\\
L. Martina\footnote{e-mail: Luigi.Martina@le.infn.it}
\\
Dipartimento di Fisica dell'Universit\`a
\\
and\\
Sezione INFN di Lecce. Via Arnesano, CP. 193\\
I-73 100 LECCE (Italy).
\\ and\\
P.~C.~Stichel\footnote{e-mail: peter@Physik.Uni-Bielefeld.DE}
\\
An der Krebskuhle 21\\
D-33 619 BIELEFELD (Germany)
}

\date{\today}

\maketitle

\begin{abstract}
    The coupling of non-relativistic anyons
    (called exotic particles)  to an
    electromagnetic field is considered. 
    Anomalous coupling is introduced by adding 
    a spin-orbit term to the Lagrangian. Alternatively, one has two
    Hamiltonian structures, obtained by either adding the anomalous 
    term to the Hamiltonian, or by redefining the mass and the NC parameter.
    The model can also be derived from its relativistic counterpart.
\end{abstract}

\vskip0.5mm\noindent

\section{Introduction}

Anyons (by which we mean here a 
 particle in the plane which carries fractional spin)
  \cite{Anyons, anyonBMT}
with anomalous gyromagnetic ratio have recently been considered
\cite{AnAn,Rapid}  either in Souriau's   \cite{SSD} symplectic,
or in a novel, ``enlarged Galilean'' framework.
Both approaches are somewhat unfamiliar to most physicists.
In this Letter we continue our investigations using
 more conventional methods, close to the spirit of Ref.
 \cite{Scripta}.

\goodbreak

\section{Exotic particles with minimal electromagnetic 
interaction}\label{nonrel}

A curious fact  known for thirty years but
 only investigated in more recent times
is that the planar Galilei group admits a
two-fold ``exotic'' central extension, labeled with 
$m$ (the mass) and a second, ``exotic'' parameter $\kappa$ 
\cite{exotic}. Physical realizations
of this symmetry have been presented, independently \cite{LSZ1,DH};
both can be obtained from their relativistic anyons  
as  ``Jackiw-Nair''  (JN) limits \cite{JaNa,HP1}. 
  The first of these models, referred to as the 
``extended exotic particle'', uses an acceleration dependent 
Lagrangian
\cite{LSZ1}.
  In terms of (external) momenta, $P_{i}$, and suitably defined
 external and internal coordinates 
$X_{i}$ and $Q_{i}$ and  ($i=1,2$), \cite{HP1,LSZ2},
the model is conveniently described by the first-order Lagrangian
\begin{equation}
    L^0=L_{ext}^0+L_{int}^0=
    \Big\{    
P_{i}\cdot\dot{X}_{i}+\frac{\theta}{2}\epsilon_{ij}P_{i}\dot{P}_{j}
    -\frac{\vP^2}{2m}
    \Big\}+\Big\{
    \frac{1}{2\theta}\epsilon_{ij}Q_{i}\dot{Q}_{j}
    +\frac{1}{2m\theta^2}\vQ^2\Big\},
    \label{freelag}
\end{equation}
where we introduced the non-commutative parameter 
$\theta=\kappa/m^2$. $\vQ^2$ is a constant of the motion.
When  $\vQ^2=0$, the internal space reduces to a point, and we 
recover the ``minimal'' exotic particle in \cite{DH}.
We first consider the extended case. $Q_{i}=0$.
The nontrivial Poisson-bracket relations are 
 \begin{eqnarray}
    \{X_{i},X_{j}\}=-\{Q_{i},Q_{j}\}=\theta\epsilon_{ij},
    \quad
     \{X_{i},P_{j}\}=\delta_{ij}.
     \label{commrel}
\end{eqnarray}


Such a particle can be coupled minimally to an electromagnetic field 
in various ways.

(i) One possibility \cite{HP1,LSZ2} is to couple to the external part 
only
by adding the usual expression  
\begin{equation}
L_{ext}^{gauge}=e(A_{i}\dot{X}_{i}+A_{0}),
\label{i}
\end{equation}   
which amounts to  gauging the global symmetry 
associated with the electric charge. 
This amounts to modifying the symplectic structure which determines 
the
non-commutative geometry of the phase space, cf. (\ref{NRPB}) below.

(ii) In another scheme \cite{LSZ2} the Hamiltonian is 
\begin{equation}
H_0=\frac{\vP{}^2}{2m}\to
\frac{1}{2m}\Big(\vP-e\vA\Big)^2-eA_0
\label{ii}
\end{equation}
while the non-commutative geometry is unchanged.
In such a way the interaction changes Abelian gauge transformations 
\cite{LSZ2}
\footnote{Yet another  coupling scheme is put forward in \cite{HP3}.}.
The two schemes are equivalent in the absence 
of the exotic structure, $\theta=0$, but not for $\theta\neq0$.

Both schemes leave the internal motions uncoupled.
They can be also coupled, however, by  gauging the  additional
``internal'' global  SO$(2)$ symmetry, 
$ 
\delta Q_{i}=\varphi\,\epsilon_{ij}Q_{j}
$ 
$\varphi\in\IR$  \cite{HP1}. In scheme (i)  the 
interaction of  an ``extended exotic particle'' with an
electromagnetic field is described by the Lagrangian
\begin{eqnarray}
    L=L^0+L_{ext}^{gauge}+L_{int}^{gauge},
    \qquad 
    L_{int}^{gauge}=\frac{\vQ^2}{2\theta}
    \big(A_{i}\dot{X}_{i}+A_{0}\big).\label{totlag}
\end{eqnarray}
Then  easy calculation shows that the Lagrangian (\ref{totlag}) is  
quasi-invariant with respect to {\it local} internal rotations 
 supplemented by a gauge transformation,
$
\delta Q_{i}=\varphi(\vX,t)\,\epsilon_{ij}Q_{j},
    \;
\delta A_{\mu}=\p_{\mu}\varphi.
$ 
 The coefficient in the interaction term
(\ref{totlag}) is fixed
by gauge invariance: it  generates internal rotations,
$ 
\left\{\vQ^2,Q_{i}\right\}=2\theta\epsilon_{ij}Q_{j}.
$ 
The Euler-Lagrange equations are
\begin{eqnarray}
    m^* \dot{X}_{i}&=&P_{i}-em\theta\epsilon_{ij}E_{j},\label{vitesse}
    \\[4pt]
    \dot{P}_{i}&=&e B\epsilon_{ij}\dot{X}_{j}+e 
    E_{i},\label{Lorentz}
    \\[4pt]
    \dot{Q}_{i}&=&\epsilon_{ij}Q_{j}    
\Big(A_{k}\dot{X}_{k}+A_{0}+\frac{1}{m\theta}\Big),\label{partrans}
    \label{eqmot}
\end{eqnarray}
where $E_{i}$ and $B$ are the electric and magnetic field, 
and $e$  denotes the shifted charge $e+\vQ^2/2\theta$.
$
m^*=m(1-e\theta B),
$ 
 is the  effective mass introduced in \cite{DH}.
Equation (\ref{partrans}) 
implies at once that the [squared] length
of the internal vector, $\vQ^2$ (and hence also the
shifted charge) are constants of the motion.
\goodbreak

 In the general case, the ``internal'' 
variable is parallel transported, just like for a 
particle with nonabelian internal structure \cite{Wong}.
This motion is, however, a mere gauge artifact that could 
be eliminated by a gauge transformation with $\varphi(t)=-t/m\theta$, 
which would also remove the $(m\theta)^{-1}$ in (\ref{partrans}).
The only physical quantity is $\vQ^2$.
Being unphysical, the motion of the internal variable $\vQ$ will, 
therefore not be considered in what follows.
We only consider the 
equations  (\ref{vitesse}-\ref{Lorentz}).

 When $\vQ=0$, we 
recover the ``minimal'' exotic particle of \cite{DH}, coupled to 
an e.m. field.

In the second scheme (ii), the electromagnetic interaction including 
the internal motion can be obtained, as described in \cite{LSZ2},
 from (\ref{ii}) by means of a noncanonical transformation of the 
phase space variables, supplemented with a
classical Seiberg-Witten map between the corresponding gauge 
potential.

Therefore, in both cases, the additional coupling to internal motion 
amounts to replacing the original, ``bare'' charge by the total 
charge, $e\to e+\vQ{}^2/2\theta$, 
whose two parts can't be measured separately. 

\goodbreak

\section{Anomalous coupling}\label{anom}

Anomalous coupling to the electromagnetic field has been
studied before \cite{AnAn,Scripta,Dixon,JMS,Kunzle, anomany}.
 The traditional rule of nonrelativistic physics,
translated into the plane, says that
magnetic moment interactions should be introduced
by adding a term $\mu B$ to the Hamiltonian, where
$ 
\mu={egs_{0}}/{2m}
$
is the magnetic moment. Here $g$ is the gyromagnetic ratio and 
we denoted non-relativistic spin by $s_{0}$.
Here we propose to generalise this rule by also including
an electric term, namely by adding to (\ref{totlag})
\begin{equation}
    L_{anom}=\mu B-\frac{g}{2}e\theta
    \epsilon_{ij}P_{i}E_{j},
    \qquad
    \mu=\frac{ge}{2m}s_{0}.
    \label{anocoup}
\end{equation}
The equations of motion look rather complicated,
\begin{eqnarray}
    m^*\dot{X}_{i}&=&P_{i}-
    \Big(1-\frac{g}{2}\Big)em\theta\epsilon_{ij}E_{j}
    -\mu m\theta\epsilon_{ij}\p_{j}B+
    \frac{emg\theta^2}{2}\left(P_{i}\p_{k}E_{k}-
    P_{k}\p_{k}E_{i}\right),
    \label{anomvitesse}
    \\[4pt]
    \dot{P}_{i}&=&e\big(E_{i}+B\epsilon_{ij}\dot{X}_{j}\big)+
    \mu\p_{i}B-\frac{eg\theta}{2}\epsilon_{kj}P_{k}\p_{i}E_{j}.
    \label{anomLorentz}
\end{eqnarray}

$\bullet$ for $g=0$ we plainly recover the previous equations of 
motion (\ref{vitesse}-\ref{Lorentz}-\ref{partrans}).

$\bullet$ By (\ref{anomvitesse}) the velocity 
and the momentum, $\dot{X}_{i}$ and $P_{i}$, respectively,
are not parallel in general, except for 
$g=2$ and for constant magnetic and linear and central electric field.

$\bullet$ When the   fields are not only weak but also {\it 
constant}, eqns. 
(\ref{anomvitesse}-\ref{anomLorentz}) reduce to the
weak-field, non-relativistic equations, \# (7.1) of \cite{AnAn}, i. e.,
\begin{equation}
      \begin{array}{ll}
	 m^\star\,\dot{X}_i= P_{i}
     -\left(1-\displaystyle\frac{g}{2}\right)m\theta\epsilon_{ij}eE_{j},
	\\[12pt]
	\dot{P}_{i}= e\big(E^{i}+B\epsilon_{ij}\dot{X}_j\big)
	\end{array}
\label{DHeqmot}
\end{equation}
These equations are Hamiltonian. The commutation relations are those 
of an ``ordinary'' exotic particle, \cite{DH}, and  
the spin-orbit term is added to the Hamiltonian,
\begin{eqnarray}
    \{X_{i},X_{j}\}=\frac{\theta}{1-e\theta B}\,\epsilon_{ij},
    \quad
    \{X_{i},P_{j}\}=\frac{1}{1-e\theta B}\,\delta_{ij},
    \quad
    \{P_{i},P_{j}\}=\frac{eB}{1-e\theta B}\,\epsilon_{ij}\label{NRPB}
    \\[6pt]
    \widetilde{H}=\left(\frac{\vP{}^2}{2m}+A_0+\mu B\right)
    +\frac{g}{2}\,e\theta\epsilon_{ij}P_iE_j.\label{NRHam}
\end{eqnarray}
($B\neq B_c$) where $A_0=-eE_iX_i$.
The generic solutions of the equations of motion (\ref{DHeqmot}) are of the 
familiar cycloidal form describing the Hall drift of the guiding center 
combined with uniform rotations with frequency
\begin{equation}
     \Omega=\frac{eB}{m^*}.
     \label{freq}
\end{equation}
Unlike in \cite{Rapid},
the ``corrected" Larmor frequency only depends on the 
non-commutative parameter $\theta$ but is independent of
 the gyromagnetic ratio $g$.

Remarkably, the same equations (\ref{DHeqmot}) can be derived also 
from another Hamiltonian structure, namely from
\begin{eqnarray}
    \Big\{\!\Big\{X_{i},X_{j}\Big\}\!\Big\}&=&
    \displaystyle\frac{1-(g/2)}{1-e\theta B}\,
    \theta\,\epsilon_{ij},\label{DHXX}
    \\[6pt]
    \Big\{\!\Big\{X_{i},P_{j}\Big\}\!\Big\}&=&
    \displaystyle\frac{1-(g/2)e\theta B}{1-e\theta B}\,\delta_{ij},
    \label{DHXP}
    \\[8pt]
    \Big\{\!\Big\{P_{i},P_{j}\Big\}\!\Big\}&=&
    \displaystyle\frac{1-(g/2)e\theta B}
    {1-e\theta B}\,eB\,\epsilon_{ij}\label{DHPP}
     \\[6pt]
    H&=&\displaystyle\frac{\vP{}^2}{2m\big(1-(g/2)e\theta B\big)}
    +A_0+\mu B.
    \label{DHHam}
\end{eqnarray}
These are indeed the usual ``exotic'' relations, but with redefined 
NC parameter and mass,
\begin{equation}
\theta\to\frac{1-(g/2)}{1-(g/2)e\theta B}\,\theta
\qquad
m\to m\big(1-(g/2)e\theta B\big),
\end{equation}
respectively. Thus, for constant external fields,
the anomalous electric coupling term in (\ref{anocoup}) (or (\ref{NRHam}))
can be suppressed by redefining the parameters,
yielding the same  equations (\ref{vitesse}-\ref{Lorentz}) as in the minimal model. 
 The constant  term $\mu B$ can actually be dropped from both 
(\ref{NRHam}) and  (\ref{DHHam}).

\section{Relation to relativistic anyons}\label{relat}

The anomalous theory of Ref. \cite{AnAn}
was based on replacing the (relativistic)
``bare'' mass by a field-dependent expression,
$ 
m\to M=M(e\FS),
$
where $S_{\alpha\beta}$ is the spin tensor, and
$\FS=-S_{\alpha\beta}F^{\alpha\beta}$ \cite{Dixon,JMS}
\footnote{Greek indices refer to
$2+1$ dimensional Minkowski space.}.
Now in the plane the usual requirement
$S_{\alpha\beta}P^\beta=0$ implies that
 spin is given by the momentum,
\begin{equation}
S_{\alpha\beta}=\frac{s}{M}\,
\epsilon_{\alpha\beta\gamma}P^{\gamma}.
\end{equation}
In \cite{AnAn} the choice was
\begin{equation}
    {\widetilde{M}}^2=m^2+\frac{ge}{2c^2}\FS.
    \label{m2}
\end{equation}
 It should be stressed, however, that (\ref{m2})
is a mere  Ansatz, and does {\it not} follow from
any first principle. In fact, {\it any} function 
$M=M(e\FS)$ would yield a consistent theory
\cite{Scripta,Dixon,JMS}.
For example, 
\begin{equation}
    M=m+\frac{ge}{4mc^2}\FS
    \label{masstilde}
\end{equation}
could be (and has been \cite{Kunzle}) used.
In the weak-field-limit,  (\ref{masstilde}) yields the same
equations as (\ref{m2}), since $\widetilde{M}\approx M$
if $eg\FS/m^2c^2<<1$. 
In what follows, we shall use the simpler expression 
(\ref{masstilde}).
Then the procedure followed in \cite{AnAn} is readily seen to
be equivalent, in the weak-field limit,
to adding to Cartan's variational 1-form  (whose integral is the
classical action \cite{SSD}) the anomalous spin-field term 
\begin{equation} 
    \Delta\alpha=-\frac{ges}{4mM}
    \epsilon_{\alpha\beta\gamma}P^{\alpha}F_{\beta\gamma}
    \left(\frac{P_{\sigma}dX^\sigma}{Mc^2}\right).
    \label{RSO}
\end{equation}
But we can parametrize our curves with proper time,
$ 
(P_{\alpha}dX^\alpha)/Mc^2=d\tau
$ 
 \cite{AnAn}.
The extra term has, therefore, the same effect as adding 
\begin{equation}
   \Delta H=\frac{ges}{4mM}\epsilon^{\alpha\beta\gamma}
    P_{\alpha}F_{\beta\gamma}
    \label{anoRcoup}
\end{equation}
to the Hamiltonian, since
$\displaystyle\int\!\Delta\alpha=-\displaystyle\!\int\Delta Hd\tau$.
\goodbreak

In a local Lorentz frame, putting 
$ 
s=\theta m^2c^2+s_{0}
$
 allows us to infer that the extra piece added to the Lagrangian is
$$
+\frac{gem\theta B}{2M}\,P_{0}
+\frac{ges_{0}}{2m}B\left(\frac{P_{0}}{Mc^2}\right)
-\frac{g}{2}\frac{m}{M}e\theta\epsilon_{ij}P_{i}E_{j}
-\frac{1}{c^2}\,\frac{ges_{0}}{2mM}\epsilon_{ij}P_{i}E_{j}.
$$
 $P_{0}\approx Mc^2$ and $m/M\approx 1$  in the NR limit.
Removing the first, divergent term and dropping the last
one which goes to zero as $c\to\infty$. 
In the JN limit, neglecting higher-order terms, we end up
with $L_{anom}$ with $Q=0$ in (\ref{anocoup})
\footnote{The $Q\neq0$ case could be studied
starting with the ``particle with torsion'' 
\cite{torsion}.}.
Alternatively, the spin-orbit term $H_{anom}$ in (\ref{NRHam}) is the 
JN
limit of (\ref{anoRcoup}).
The two possibilities i. e., either changing the kinetic term,
 or adding a spin-orbit piece to the Hamiltonian
are the relativistic counterparts of the two Hamiltonian structures
we found in the non-relativistic context.

\section{Semiclassical Dirac particle}

Returning to the non-relativistic setting, let us illustrate our 
theory on a related problem. 
In a recent paper  \cite{BM2}, B\'erard and Mohrbach consider
 a $3D$ Dirac particle in a constant electric field and show that, 
semiclassically, the particle admits, to order 
$c^{-2}$, the anomalous velocity relation
\begin{equation}
    m\frac{dX_{i}}{dt}\approx P_{i}-
    \frac{1}{2}\,\frac{e}{mc^2}\epsilon_{ijk}\sigma_jE_{k}
    \label{BMReqmot}
\end{equation}
[supplemented with the Lorentz 
force law $\dot{P}_{i}=eE_{i}$], where $\vsigma$ is the spin vector. 
Assuming cylindrical symmetry and spin-polarized electrons,
$\sigma_{i}=-s\delta_{i3}$, 
the JN limit $s/m^{2}c^{2}\to\theta$ yields
\begin{equation}
    m\frac{dX_{i}}{dt}\approx P_{i}
    -\frac{1}{2}em\theta\epsilon_{ij}E_{j},
    \label{BMeqmot}
\end{equation}
which is the first equation in (\ref{DHeqmot}) with  $B=0$
and with anomalous gyromagnetic factor $g=1$. This value has
already been found before \cite{BD}.  
 To leading order in $c^{-1}$, the relativistic Hamiltonian behaves as
\begin{equation}
\widetilde{H}\approx mc^2+\frac{\vP^2}{2m}-e\vE\cdot\vX
    +\frac{e}{2m^2c^2}\vsigma\cdot(\vE\times\vP)
    \longrightarrow    
\frac{\vP^2}{2m}-e\vE\cdot\vX+\frac{1}{2}e\theta\epsilon_{ij}P_iE_j.
    \label{BMHam}
\end{equation}
cf. (\ref{NRHam}).
Note that the naive Hamilton equation, 
$\dot{X}_{i}=\p\widetilde{H}/\p P_{i}$,
would contain a factor $(+1/2)$ instead of  $(-1/2)$ in front of
the anomalous term in (\ref{BMeqmot}). 
The correct coefficient is recovered when the exotic part is taken
into account. Either of the Hamiltonian structures  
\begin{eqnarray}
    \{X_{i},X_{j}\}_{\alpha}=(1-\alpha)\theta \epsilon_{ij},
    \qquad
    \{X_{i},P_{j}\}_{\alpha}=\delta_{ij},
    \qquad
    \{P_{i},P_{j}\}_{\alpha}=0,
    \\[6pt]
    H_{\alpha}=
    \frac{\vP^2}{2m}-e\vE\cdot\vX+\Big(\frac{1}{2}-\alpha\Big)
    e\theta\epsilon_{ij}P_iE_j,
\end{eqnarray}
yields indeed the correct equations 
for any value of the real parameter $\alpha$.  
(\ref{NRPB})-(\ref{NRHam}) corresponds to $\alpha=0$, and  
 (\ref{DHXX})-(\ref{DHXP})-(\ref{DHPP})-(\ref{DHHam})
 corresponds to $\alpha=1/2$, respectively.
\goodbreak

\section{Further generalizations}

A  slightly modified model is obtained  
replacing the momentum in (\ref{anocoup}), $P_{i}$, 
by the velocity, $\dot{X}_{i}$~:
\begin{equation}
    L_{anom}'=\mu B-\frac{g}{2}me\theta\,
    \epsilon_{ij}\dot{X}_{i}E_{j},
    \label{JSternanocoup}
\end{equation}
Magnetic moment interaction of such kind has been considered
before \cite{anomany}. Eqn. (\ref{JSternanocoup}) is also
reminiscent of the interaction of a magnetic 
moment with an electric charge \cite{AhaCash}. 

Adding (\ref{JSternanocoup}) to our Lagrangian
(\ref{totlag}) amounts indeed to changing the potentials in 
(\ref{vitesse})-(\ref{Lorentz})-(\ref{partrans}) according to
$ 
A_{0}\to A_{0}'=A_{0}+(\mu/e)B,
$ 
$
A_{i}\to A_{i}'=A_{i}-(mg\theta/2)\epsilon_{ij}E_{j},
$ 
that yields 
$ 
B\to B'=B+mg\theta/2\p_{k}E_{k},
$ 
$E_{i}\to E_{i}'=E_{i}+{\mu}/{e}\p_{i}B
+{mg\theta}/{2}\epsilon_{ij}\p_{t}E_{j}.
$
 Eliminating the momenta in the new equations of motion
and dropping terms which contain second derivatives of
the field, we obtain
\begin{equation}
    \frac{\ d}{dt}\Big({m^*}'\dot{X}_{i}\Big)=
    e(B\epsilon_{ij}\dot{X}_{j}+E_{i})+\mu\p_{i}B
    -me\theta\epsilon_{ij}\frac{dE_{j}}{dt}
    +\frac{emg\theta}{2}\epsilon_{ij}(\dot{X}_{j}\p_{k}E_{k}
    +\p_{t}E_{j})
    \label{modNewton}
\end{equation}
with the new magnetic field, $B'$, replacing $B$ in the 
new effective mass,
$
m\to{m^*}'=m(1-e\theta B').
$  
 For the sake of comparision, neglecting terms which are higher-order 
in the fields, from (\ref{anomvitesse}-\ref{anomLorentz}),
we would get instead
$$ 
    \frac{\ d}{dt}\Big({m^*}\dot{X}_{i}\Big)=
    e(B\epsilon_{ij}\dot{X}_{j}+E_{i})+\mu\p_{i}B
    -me\theta\epsilon_{ij}\frac{dE_{j}}{dt}
    +\frac{emg\theta}{2}\Big(
    \epsilon_{ij}\dot{X}_{k}\p_{k}+\p_{t}
    -\epsilon_{kj}\dot{X}_{k}\p_{i}\Big)E_{j}.
$$ 
This is readily transformed into the form (\ref{modNewton}). 
In a weak and slowly varying field, the two models
 only differ in the form of the effective mass.
 
It is worth remembering that anomalous velocity 
relations of the type studied here have been considered in
the context of the Anomalous Hall Effect \cite{AHE} and in the
semiclassical theory of the Bloch electron \cite{Niu}.
Equations (\ref{vitesse})-(\ref{Lorentz}), 
or their ``anomalous" generalization in constant
external fields, (\ref{DHeqmot}), is indeed a special  
case of the more general system
\begin{eqnarray}
  \dot{X}_{i} +
    \theta(\vP) \epsilon_{i j}\dot{P}_j&=&\p_{P_{i}}{\cal E},  
    \label{EMcoupling1}
    \\
    eB\epsilon_{i j} \dot{X}_j-\dot{P}_{i}&=&-e E_i,
     \label{EMcoupling2}
\end{eqnarray}
where ${\cal E}={\cal E}_0(\vP)-B {\cal M}(\vP)$ 
is the total energy with ${\cal E}_0$ and ${\cal M}$
denoting the Bloch band energy and the magnetization, respectively. 
 These equations  can
be derived, under quite general
assumptions, by semiclassical calculations applied to the dynamics of 
wave packets in a two-dimensional  crystal  \cite{Niu}.
Note that the non-commutative parameter has been promoted to
a function of the momentum \cite{BM}.

 The system (\ref{EMcoupling1}-\ref{EMcoupling2}) can actually be reduced to 
first order equations for the $P_i$ alone, 
\begin{equation}
\left(1-eB\theta(\vP)\right)\dot{P}_{i}
    = e B\epsilon_{ij}\p_{P_{j}}{\cal E}+eE_i,
     \label{reduced}
\end{equation}
that can be integrated 
by solving with respect to $P_1$, say,
using the conserved quantity 
\begin{equation}
{\cal C}= {\cal E}-\frac{\epsilon_{ij}P_i E_j}{B}.
     \label{conserpspace}
\end{equation}
 Thus the problem is reduced to quadratures. Note that eqn. 
 (\ref{reduced}) is actually Hamilton's equation for ${\cal C}$
as Hamiltonian and Poisson bracket (\ref{NRPB}c) in $P$-space
alone.

In conclusion, we mention that another way of introducing anomalous coupling  
for constant e.m. fields has been advocated by us in \cite{Rapid}. 
There we introduced an ``enlarged'' planar Galilei group, which 
incorporates field variables besides space-time.  
Interestingly, the square of (\ref{conserpspace}) is proportional
to a Casimir of the enlarged symmetry algebra in \cite{Rapid}, and 
anomalous coupling can then be achieved by adding this Casimir  
to the Hamiltonian.

\kikezd{Acknowledgement}.
PAH and PCS would like to thank for hospitality
the University of Lecce. 



\end{document}